\documentclass[aps,preprint,a4paper,12pt]{revtex4-1}
\usepackage[utf8]{inputenc}
\usepackage{amsmath}
\usepackage{hyperref}
\usepackage{amsfonts}
\usepackage{amssymb}
\usepackage{graphicx}
\usepackage{latexsym}
\usepackage{color}
\usepackage{booktabs}
\usepackage{dcolumn}
\usepackage{epsfig}
\usepackage{subfigure}
\usepackage{float}
\usepackage{multirow}
\usepackage{ulem}
\usepackage{xcolor}
\usepackage{svg}
\usepackage{tikz}

\def\be{\begin{eqnarray}}
\def\ee{\end{eqnarray}}

\begin{document}

\title{Gauss-Bonnet black holes in $(2+1)$ dimensions: Perturbative
  aspects and entropy features}

\author{B. Cuadros-Melgar}
\email{bertha@usp.br}
\affiliation{Escola de Engenharia de Lorena, Universidade de S\~ao
   Paulo, Estrada Municipal do Campinho S/N, CEP 12602-810, Lorena, SP, Brazil}

\author{R. D. B. Fontana}
\email{rodrigo.dalbosco@ufrgs.br}
\affiliation{Universidade Federal do Rio Grande do Sul, Campus Tramanda\'i-RS \\
Estrada Tramanda\'{i}-Os\'orio, CEP 95590-000, RS, Brazil}

 \author{Jeferson de Oliveira}\email{jeferson.oliveira@ufmt.br}
 \affiliation{Instituto de F\'i­sica, Universidade Federal de Mato Grosso, CEP 78060-900, Cuiab\'a, MT, Brazil}

\date{\today}

%------------------------------------------------------------%
\begin{abstract}
We investigate some aspects of the $(2+1)$-dimensional Gauss-Bonnet black hole
proposed in~\cite{Hennigar:2020fkv,Hennigar:2020drx}. The
perturbations of scalar and massless spinorial fields are studied
suggesting the dynamical stability of the geometry. The field
evolution is analyzed calculating the quasinormal modes for different
parameters and exploring the influence of the coupling constant of the
theory. The hydrodynamical modes are also obtained in the small
coupling limit. Furthermore, the entropy bound and the dominant
semiclassical correction to the black hole entropy are calculated. 
\end{abstract}
%------------------------------------------------------------%

\maketitle

%------------------------------------------------------------%

\section{Introduction}

Lower dimensional gravity has been a very active field for a long time
in theoretical physics due to both its simplicity and its features,
which have a strong similarity to those in the four-dimensional
gravity theory. Black hole solutions were found in several lower
dimensional models like the well-known $(2+1)$-dimensional
BTZ ({\it{Bañados-Teitelboim-Zanelli}}) black
hole~\cite{Banados:1992wn} and the solutions of Jackiw
gravity~\cite{Jackiw:1984je} in $(1+1)$-dimensions among others (for an
extensive survey see~\cite{Mann:1991qp}). The Gauss-Bonnet (GB)
gravity is a particular case of Lovelock
theories~\cite{Lovelock:1971yv}, which includes higher curvature
corrections to the Einstein-Hilbert action given in terms of the
Riemann tensor. The equations of motion are differential equations of
second-order for the metric tensor components. A special feature of
those higher-curvature terms is that they are identically zero if the
spacetime dimension is bounded by $D<5$.  

More recently, a proposal to evade the Lovelock theorem and allow
higher-curvature terms, in particular, the GB term, to survive without
extra fields for $D < 5$ was proposed
in~\cite{Glavan:2019inb}. Nevertheless, in several papers its was shown 
that such a proposal leads to an ill-defined
theory~\cite{Gurses:2020ofy,Hennigar:2020lsl,Arrechea:2020evj}. Despite 
such inconsonance, it is still possible to include the
four-dimensional GB corrections in a consistent
way~\cite{Lu:2020iav,Kobayashi:2020wqy}, showing that the 
four-dimensional solution reported in~\cite{Glavan:2019inb} could be
obtained from a scalar-tensor theory which is a subclass of Horndeski
family~\cite{Horndeski:1974wa}. Following the same guidelines, a
$(2+1)$-dimensional black hole solution with GB correction was found
by Henningar {\it{et
    al.}}~\cite{Hennigar:2020fkv,Hennigar:2020drx}. Such a
family of solutions admits a generalization of the BTZ black hole, which is
recovered in the limit when the GB coupling goes to zero. In our
present work we are interested in a deeper comprehension of those GB-BTZ
black holes in $(2+1)$ dimensions, specially in the role that the GB coupling
constant plays on the stability when the metric is perturbed by probe fields.

As it is essential to understand in which situations a black hole
solution is stable under small perturbations, the study of the field
propagation and the determination of the quasinormal spectrum
due to probe matter fields in the geometry of the (2+1)-dimensional
GB-BTZ black holes can shed some light on this stability. Moreover,
the stable or unstable nature of the metric is closely related to
the shape of each wave potential~\cite{hor2000}.

Much work has been done on linear perturbations of GB black holes in
different dimensions. Recent studies include the gravitational
perturbations and the ringdown phase of black holes in
Einstein-dilaton-Gauss-Bonnet gravity in four
dimensions~\cite{Blazquez-Salcedo:2016enn} and the 
use of quasi-periodic oscillations to constrain the space of
parameters of the theory~\cite{Maselli:2014fca}. In addition, 
in~\cite{Konoplya:2019fpy} the calculation of the black hole shadow
radius is implemented in the Einstein-scalar-Gauss-Bonnet gravity with
non-trivial scalar hair and in~\cite{Konoplya:2020bxa} the quasinormal
modes and the stability of the new four-dimensional Gauss-Bonnet black
holes were investigated. Moreover, an interesting relation between the
shadow radius and quasinormal spectrum was stablished
in~\cite{Jusufi_2020a,Jusufi_2020,Cuadros_Melgar_2020}. Whether such a 
relation can be assigned to three-dimensional black holes still
remains an open question.   

In addition, we are interested in exploring some thermodynamical aspects
of the (2+1)-dimensional GB-BTZ black hole. Since the pioneering works of
Bekenstein~\cite{PhysRevD.7.949} and Hawking~\cite{Hawking1975}, which
led to the identification of the black hole surface gravity and the
event horizon area with the temperature and the entropy of a
thermodynamical system, respectively, the black hole thermodynamics has
developed and brought different techniques that have improved our
understanding of the properties of these remarkable objects. One of
these physical quantities is the black hole entropy, which accounts for
the maximum entropy a physical system can carry. If an object is
captured by a black hole, according to the generalized second law of
thermodynamics, the entropy should always increase as well as the
event horizon area since they are connected through the
Bekenstein-Hawking classical formula, $S_{BH} = Area/4$. Based on this
observation, Bekenstein proposed the existence of an upper bound on
the entropy of any system~\cite{PhysRevD.23.287} carrying an energy
$E$ and with a characteristic dimension $R$, {\it i.e.}, $S \leq 2\pi ER$, which
proved to be universal until nowadays. Along with this subject is the
need to include quantum aspects in the description of black hole
entropy. In this way, 't Hooft brought a proposal forward by considering
a thermal bath of scalar fields just outside the event horizon so that
they could contribute to the entropy provided that a cut-off both close and
far from the black hole is included. This technique is known as the
brickwall method~\cite{THOOFT1985727} and its calculation leads to a
dominant correction also correlated to the event horizon area. In
fact, the coefficient of proportionality is universal for each
spacetime dimensionality. It is our aim to verify if these properties
can be fulfilled by the (2+1)-dimensional GB-BTZ black hole. 

The paper is organized as follows. In Sec.\ref{sec1} we discuss the
main features of $(2+1)$ GB-BTZ black hole solution. In Sec.\ref{sec2}
we compute the quasinormal modes and frequencies due to a massless
scalar probe field and discuss the effect of the GB coupling upon the
stability. Sec.\ref{sec3} brings the massless spinorial field as the
probe field and the quasinormal modes and spectrum are obtained. In
Sec.\ref{sec4} the hydrodynamic approximation for the probe scalar
field in the limit of small GB coupling constant is considered and its
interpretation in terms of gauge/gravity correspondence is
discussed. Sec.\ref{sec5} is devoted to some thermodynamical aspects
of the black hole solutions. Finally, in Sec.\ref{sec6}, we discuss
our results and possible perspectives for future work.

%------------------------------------------------------------%
\section{Gauss-Bonnet black hole solutions in $(2+1)$-dimensions}\label{sec1}

The action that describes the Gauss-Bonnet (GB) gravity in
$(2+1)$-dimensions, encoding the main characteristics of its
$(3+1)$-dimensional counterpart, is given by \cite{Hennigar:2020fkv}, 

\begin{equation}\label{action}
S = \int d^3 x \sqrt{-g}\left\{R-2\Lambda + \alpha\left[\phi
  \mathcal{G} +4G^{ab}\partial_{a}\phi\partial_{b}\phi -
  4(\partial\phi)^{2}\square\phi +
  2((\nabla\phi)^2)^{2}\right]\right\}\,, 
\end{equation}
where we have the Einstein-Hilbert term plus a cosmological constant
$\Lambda$, the corrections coming from the GB term~\footnote{Notice
  that the GB term identically vanishes in a
  $(2+1)$-dimensional spacetime.}
$\mathcal{G} = R_{abcd}R^{abcd} - 4R_{ab}R^{ab}+R^2$, being $\alpha$ the GB
coupling constant, and an additional scalar field $\phi$. Notice that
the same coupling between the Einstein tensor $G_{ab}$ and the kinetic
term of $\phi$ is present in the Horndeski theory and, indeed, the
theory represented by the action (\ref{action}) is a special case of
Horndeski class~\cite{Horndeski:1974wa}. 

As pointed out in~\cite{Hennigar:2020fkv}, the GB part of
(\ref{action}) can be obtained at least by two different methods. 
Namely, a Kaluza-Klein (KK) dimensional reduction of a $D$-dimensional theory
compactified on an internal maximally symmetric space that leads to a
$D=3$ GB gravity~\cite{Lu:2020iav} and the generalization of Ross-Mann
method to obtain the $D\rightarrow 2$ limit of General
Relativity~\cite{Mann:1992ar} through a conformal transformation on
the metric $\tilde{g_{ab}} = e^{\Psi}g_{ab}$ and an expansion of the action
around the spatial dimension of interest. Both methods lead to the
action (\ref{action}) as long as the maximally symmetric space used in
the KK approach is flat, otherwise, additional terms are
generated~\cite{Hennigar:2020fkv},  
\begin{equation}\label{action_internal}
S_{\lambda} = -2\int d^{3}x\sqrt{-g}\left[\lambda e^{-2\phi}\left(R + 6 (\partial\phi)^{2}\right) + 3\lambda^2 e^{-4\phi}\right]\,,
\end{equation}
where $\lambda$ represents the curvature of the internal space.  

In order to obtain black hole solutions to the GB gravity in
$(2+1)$ dimensions we consider the equations of motion that
come from the action (\ref{action}) together with the additional terms 
(\ref{action_internal}) and the following  {\it{ansatz}} for
the line element~\cite{Hennigar:2020fkv}, 
\begin{equation}\label{ansatz}
ds^{2} = -f(r)dt^2 + \frac{1}{f(r)h(r)}dr^2 + r^2\left(d\varphi -
\frac{J}{2r^2}dt\right)^2\,, 
\end{equation}
where $J$ is a constant. In addition, the scalar field $\phi$ depends
only on the radial coordinate, $\phi=\phi(r)$. Then, using this
{\it{ansatz}} and varying the action with respect to $f(r)$,
$h(r)$, and $\phi(r)$ we obtain three equations of motion, whose
simplest solution is the BTZ black hole~\cite{Banados:1992wn} when
considering $h(r)=1$, $\phi = \hbox{constant}$ and $\lambda = 0$, 
\begin{equation}\label{btz}
f_{BTZ}(r) = -M + \frac{r^2}{L^2} +\frac{J}{4r^2}\,,
\end{equation}
with $M$ and $J$ denoting the black hole mass and angular momentum,
respectively, and the cosmological constant $\Lambda$ is related to
the curvature radius $L$ by $\Lambda = L^{-2}$.  

Furthermore, new black hole solutions in three dimensions depending on
the GB coupling are achieved by considering a non-constant scalar
field $\phi(r)$. In the static case $J=0$ and setting $\lambda=0$ the
equations of motion admit the following
solution~\cite{Hennigar:2020fkv}, 
\begin{equation}\label{new_bh}
f(r)_{\pm} = -\frac{r^2}{2\alpha}\left(1\pm\sqrt{1+\frac{4\alpha}{r^2}\left(-M + \frac{r^2}{L^2}\right)}\right)\,, 
\end{equation}
\begin{equation}\label{new_scalar}
\phi(r) = \ln\left(\frac{r}{L}\right)\,.
\end{equation}
The positive branch $f(r)_{+}$ of solution (\ref{new_bh}) does not
have a well-defined limit as the GB coupling constant goes to
zero, in fact, in this limit it reduces to
\begin{equation}\label{positive_branch}
f(r)_{+} \approx M -\frac{r^2}{L^2} - \frac{r^2}{\alpha}\,,
\end{equation}
which goes to infinity as $\alpha\rightarrow 0$. In this sense, the
positive branch does not describe a physical system. Conversely, the
negative branch $f(r)_{-}$ reduces to BTZ black hole in the same
limit. Also, at large distances the negative branch is described by an
AdS-like metric, what yields a condition on the allowed values
of the GB coupling in order to have a well-defined solution at
spatial infinity, {\it i.e.}, $\alpha > -L^2/4$. 

Since the negative branch admits a bounded limit for small $\alpha$
and is well behaved  at large distances, we are going to consider only
$f(r)_{-}$ as black hole solution and, thus, we will drop the subscript
$-$ in $f(r)_{-}$ from now on. As the event horizon $r=r_{+}$ of this
metric is the same as that of the BTZ solution, $r_{+} = LM^{1/2}$, we see
that the GB coupling does not change the location of the event
horizon. Moreover, the near horizon limit of $f(r)$ is given by 
\begin{equation}\label{nh_metric}
f(r)\approx \frac{2M^{1/2}}{L}(r-r_{+}) + \mathcal{O}((r-r_{+})^{2})\,,
\end{equation}
showing that $\alpha$ contributes only for large distances from $r_{+}$.

Furthermore, we can distinguish two cases in the internal geometric
structure of the negative branch. When $\alpha>0$, the black hole has
a branch singularity analogous to GB higher-dimensional
solutions. This singularity can be found by using the condition that
the argument of the square root in the metric vanishes, thus we have
\begin{equation}\label{branch}
r_b = 2L \sqrt{\frac{M\alpha}{L^2+4\alpha}} = 2
\sqrt{\frac{\alpha}{L^2+4\alpha}} \,r_+ < r_+\,,
\end{equation}
where the last inequality shows that the branch singularity remains
inside the event horizon. Around $r_b$ the Kretschmann scalar behaves
as
\begin{equation}\label{Kb}
K \equiv R_{\alpha\beta\mu\nu} R^{\alpha\beta\mu\nu}\sim \frac{r_b ^3
  (L^2+4\alpha)}{32 \alpha^2 L^2 (r-r_b)^3} + \cdots \,.
\end{equation}
This type of divergence is the same found in higher-dimensional GB
solutions~\cite{PhysRevD.71.124002} and shows that $r_b$ is a true
curvature singularity. 

On the other hand, when $-L^2/4<\alpha<0$, the metric continues until
$r=0$, where the Kretschmann scalar behaves as 
\begin{equation}\label{K0}
K \sim -\frac{2M}{\alpha r^2} + \cdots \,,
\end{equation}
showing that at $r=0$ a curvature singularity is located.

The Kruskal-Szekeres extension of black solution (\ref{new_bh}) and its
Penrose-Carter diagram can be constructed by a detailed examination of
the metric near the event horizon $r=r_{+}$ and at spatial infinity
$r\rightarrow \infty$. 

Near the event horizon it is possible to approximate the function $f(r)$ as $f(r)\approx 2\kappa_{+}(r-r_{+})$, where $\kappa_{+}=f'(r_{+})/2$, and in this region the tortoise coordinate $r_{*}$ can be written as
\begin{equation}\label{tortoise_horizonte}
r_{*} \approx \frac{1}{\kappa_{+}}\ln{|\kappa_{+}(r-r_{+})|}.
\end{equation}
Defining a double null system of coordinates, $U_{+} = t - r_{*}$ and $V_+ = t + r_{*}$, we obtain the Kruskal-Szekeres extension near the event horizon,
\begin{equation}\label{KS_horizon}
U_{+}V_{+}=\mp \kappa_{+}|(r-r_{+})|,
\end{equation}
in which the minus sign refers to the region $r>r_{+}$ and the plus sign corresponds to the region $r<r_{+}$. 

At spatial infinity $r\rightarrow\infty$ the Kruskal-Szekeres extension reads
\begin{equation}\label{KS_infinito}
U_{\infty}V_{\infty} = -e^{\frac{2}{L^2 r^2}}.
\end{equation}

\begin{figure}[htb]
\begin{center}
\includegraphics[height=7.0cm, width=7.0cm]{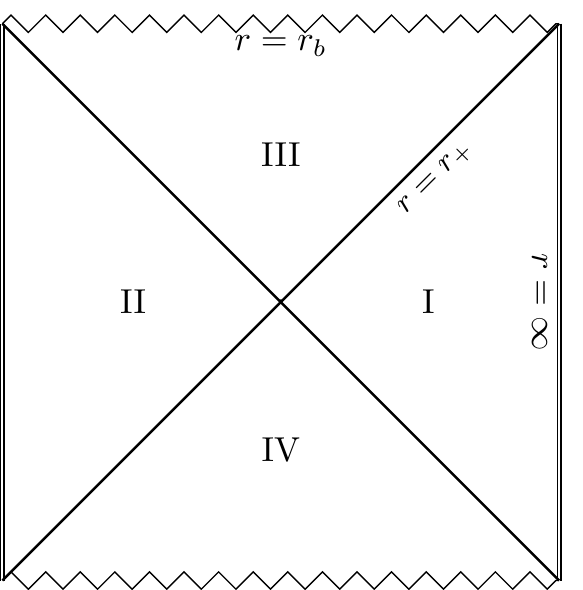}
\includegraphics[height=7.0cm, width=7.0cm]{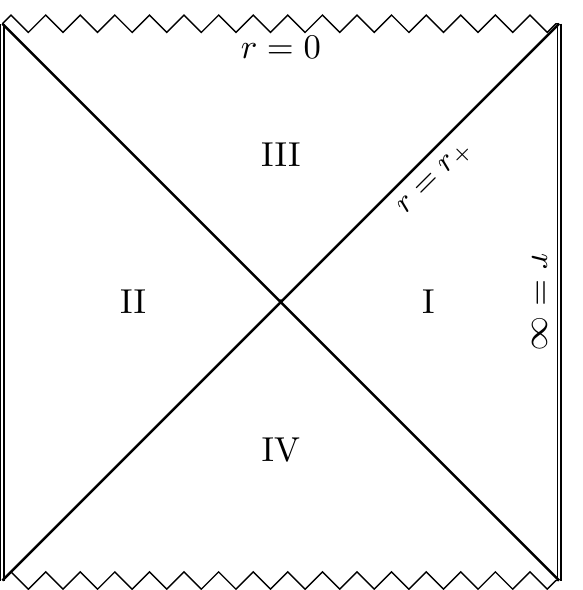}
\caption{Penrose-Carter diagrams for the $(2+1)$-dimensional
  GB-BTZ black hole with $\alpha>0$ (left) and $-L^2/4<\alpha<0$
(right).}
\label{diagrama}
\end{center}
\end{figure}

Combining each extension (\ref{KS_horizon})-(\ref{KS_infinito})
through the Penrose coordinates $T=\frac{1}{2}(\tilde{U}+\tilde{V})$
and $R=\frac{1}{2}(\tilde{U}-\tilde{V})$ with $\tilde{U}=\arctan(U)$
and $\tilde{V}=\arctan(V)$, we accomplish the Penrose-Carter diagrams
for the entire spacetime as shown in Fig. \ref{diagrama}. Notice
that the structure of these diagrams is
the same as that of the $(2+1)$-dimensional black hole in the presence
of anisotropic fluids~\cite{deOliveira:2018weu}. The spatial infinity
is conformally AdS and the nature of the singularity located at
$r=r_b$ ($\alpha>0$) or $r=0$ ($-L^2/4<\alpha<0$) is spacelike. In
both cases the singularity is always covered by an event horizon at $r=r_+$.  
%------------------%
%------------------%

After describing the main features of the black hole spacetime,
in the next sections we are going to consider two different kinds
of probe fields evolving in such geometry, namely, the massless scalar
and the massless spinor fields. The analysis of the dynamics of the
fields provides some insight on the black hole stability through
the computation of quasinormal frequencies as we will see.

%------------------------------------------------------------%
\section{Probe scalar field}\label{sec2}

 Let us consider a massless scalar field $\Psi(x^{\mu})$, whose
 dynamics is governed by the Klein-Gordon equation, 
\begin{equation}\label{kg1}
\frac{1}{\sqrt{-g}}\partial_{\mu}\left(\sqrt{-g}g^{\mu\nu}\partial_{\nu}\Psi\right)=0,
\end{equation}
in the geometry of a GB-BTZ black hole (\ref{new_bh}) with
$x^{\mu}=(t,r,\varphi)$. The tortoise coordinate defined through
$dr_{*}=dr/f$ has its domain on the region I of the diagram 
shown in Fig. \ref{diagrama}, running from $-\infty$ to a constant
value as $r \in [r_+, \infty]$. Performing the following separation of
variables  
\begin{equation}\label{separacao}
\Psi(t,r,\varphi) = \sum_{m}\frac{\psi (r,t)}{\sqrt{r}}e^{im\varphi}= \sum_{m}\frac{R(r)}{\sqrt{r}}e^{-i\omega t + im\varphi}, 
\end{equation}
the field equation (\ref{kg1}) can be cast to the form 
\begin{equation}\label{kg2}
\frac{d^2R}{dr_{*}^2} +\left(\omega^2 - V(r)\right)R = 0,
\end{equation}
in which $V(r)$ is the effective potential for the scalar field dynamics
in the black hole geometry. Explicitly, we have 
\begin{equation}\label{potential_scalar}
V(r) = f(r)\left(\frac{m^2}{r^2}-\frac{f(r)}{4r^2}+\frac{1}{2r}\frac{df(r)}{dr}\right).
\end{equation}

The effective potential $V(r)$ depends on all the parameters that
characterize the black hole geometry $(M, L, \alpha)$ and on the
scalar field azimutal number $m$.  

\begin{figure}[htb]
\includegraphics[height=10.0cm, width=15.0cm]{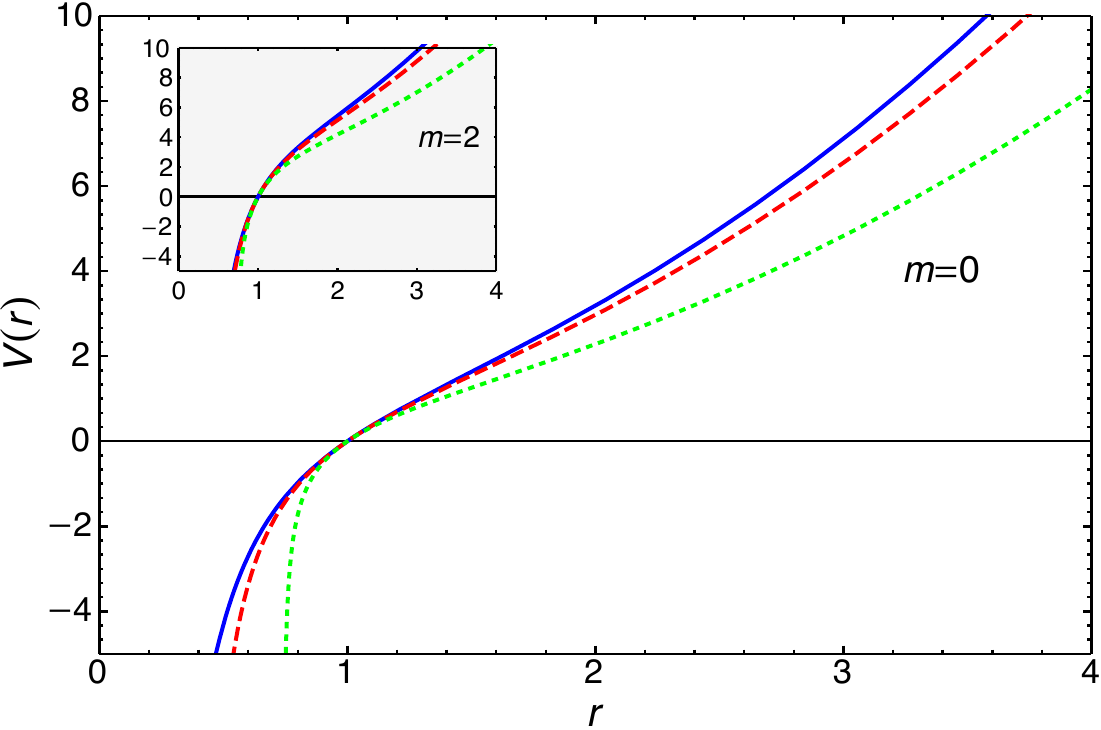} 
\caption{(color online) Main panel: Effective scalar potential $V(r)$
  with $m=0$, $M=L=1$ for different values of GB coupling
  $\alpha = 0$ (blue), $\alpha=5\times 10^{-2}$ (dashed red), and
  $\alpha = 3\times 10^{-1}$ (dotted green). Upper left panel:
  Effective scalar potential $V(r)$ with $m=2$, $M=L=1$ for different
  values of GB coupling $\alpha = 0$ (blue), $\alpha=5\times
  10^{-2}$ (dashed red), and $\alpha = 3\times 10^{-1}$ (dotted
  green).}
\label{pot_escalar}
\end{figure}

In Fig. \ref{pot_escalar} we plot different potentials varying the
GB parameter $\alpha$ with fixed $M$, $L$, and $m$. For $\alpha=0$ we
recover the effective potential for the BTZ black
hole~\cite{Cardoso:2001hn} and as $\alpha$ increases, the value of the 
potential for a given radial position $r$ decreases, showing that the
GB coupling attenuates the interaction
between the geometry and the probe massless scalar field. 
%------------------%
%------------------%

The quasinormal spectra due to the evolution of a massless scalar
field can be obtained with several known methods. We consider the
characteristic integration in double null coordinates given by
$du=dt-dr_*$ and $dv=dt+dr_*$, turning Eq. (\ref{kg2}) to the form 
\be
\label{e1}
\left(4\frac{\partial^2}{\partial u \partial v} + V(r) \right)\psi = 0.
\ee
Now, the usual discretization scheme (described in the specific
literature~\cite{Konoplya:2011qq} and references therein) gives the
following equation,  
\be
\label{e2}
\psi_N = \left( 1+ \frac{h^2}{16}V_S \right)^{-1} \left( \psi_W+\psi_E-\psi_S - \frac{h^2}{16}(V_W\psi_W+V_E\psi_E+V_S\psi_S)\right)\,,
\ee
which we can integrate yielding the field evolution with the quasinormal
signal present. After getting the field profile, acquired
through the characteristic integration, we can apply the Prony method
described in~\cite{Konoplya:2011qq} to pick up the
frequencies. 

In order to cross-check the results of the obtained quasinormal modes,
we also developed a Frobenius method, similar to that of
Ref. \cite{hor2000}. The equations for this numerical implementation
are given in Appendix \ref{ap1}. We obtained a good agreement between the
results collected with both methods for $\alpha < 0.13$. The maximum
deviation between the results of both methods appears in the case of very
small $\alpha$. Actually, for $\alpha = 10^{-4}$ we obtained an outcome with a
$2\%$ maximal deviation in the cases of higher $r_+$. Except for those
occurrences, the convergence of both methods is higher than
$0.2\%$. The Tables \ref{tb1} and \ref{tb2} display the quasinormal
frequencies for different geometry parameters. 

\begin{table}
  \centering
 \caption{The fundamental quasinormal modes for a massless scalar
   field with $L=1$ and azimutal number $m =0$.} 
\addtolength\tabcolsep{10pt}
    \begin{tabular}{ccccccc}
    \hline
	 & \multicolumn{2}{c}{$r_+=1$} & \multicolumn{2}{c}{$r_+=10$} & \multicolumn{2}{c}{$r_+=100$}	\\
	$\alpha$ & {$\Re(\omega)$} & {$-\Im(\omega)$} & {$\Re(\omega)$} & {$-\Im(\omega)$} & {$\Re(\omega)$} & {$-\Im(\omega)$} \\
	\hline \hline
$1 \cdot 10^{-4}$ &	0.0201015	&	1.99975	&	0.204054	&	19.99609	&	2.04054		&	199.96092 \\
$1 \cdot 10^{-3}$ &	0.0633286	&	1.99781	&	0.633286	&	19.97806	&	6.33286		&	199.78061 \\
$1 \cdot 10^{-2}$ &	0.198807	&	1.97948	&	1.98807		&	19.79481	&	19.88068	&	197.94808 \\
$1 \cdot 10^{-1}$ &	0.572319	&	1.79642	&	5.72319		&	17.96419	&	57.23187	&	179.64185 \\
$2 \cdot 10^{-1}$ &	0.718720	&	1.62834	&	7.18720		&	16.28342	&	71.87195	&	162.83424 \\
$3 \cdot 10^{-1}$ &	0.788886	&	1.49642	&	7.88886		&	14.96417	&	78.88860	&	149.64175 \\
$4 \cdot 10^{-1}$ &	0.825547	&	1.39082	&	8.25547		&	13.90820	&	82.55466	&	139.08200 \\
$5 \cdot 10^{-1}$ &	0.844776	&	1.30425	&	8.44776		&	13.04249	&	84.47757	&	130.42490 \\
    \hline  
    \end{tabular}
  \label{tb1}
\end{table}

\begin{table}
  \centering
 \caption{The fundamental quasinormal modes for a massless scalar
   field with $L=1$ and azimutal number $m =1$.} 
\addtolength\tabcolsep{10pt}
    \begin{tabular}{ccccccc}
    \hline
	 & \multicolumn{2}{c}{$r_+=1$} & \multicolumn{2}{c}{$r_+=10$} & \multicolumn{2}{c}{$r_+=100$}	\\
	$\alpha$ & {$\Re(\omega)$} & {$-\Im(\omega)$} & {$\Re(\omega)$} & {$-\Im(\omega)$} & {$\Re(\omega)$} & {$-\Im(\omega)$} \\
	\hline \hline
$1 \cdot 10^{-4}$	&	0.999652	&	1.99883	&	1.02051	&	19.99608	&	2.27236		&	199.96092 \\
$1 \cdot 10^{-3}$	&	1.00185		&	1.99727	&	1.18394	&	19.97796	&	6.41138		&	199.78060 \\
$1 \cdot 10^{-2}$	&	1.01893		&	1.97500	&	2.22700	&	19.79394	&	19.90599	&	197.94799 \\
$1 \cdot 10^{-1}$	&	1.13028		&	1.77622	&	5.80817	&	17.96072	&	57.24044	&	179.64150 \\
$2 \cdot 10^{-1}$	&	1.18365		&	1.60429	&	7.25065	&	16.27948	&	71.87833	&	162.83384 \\
$3 \cdot 10^{-1}$	&	1.20435		&	1.47060	&	7.94319	&	14.96002	&	78.89406	&	149.64133 \\
$4 \cdot 10^{-1}$	&	1.20937		&	1.36369	&	8.30458	&	13.90386	&	82.55958	&	139.08156 \\
$5 \cdot 10^{-1}$	&	1.20626		&	1.27595	&	8.49345	&	13.03798	&	84.48215	&	130.42445 \\
    \hline  
    \end{tabular}
  \label{tb2}
\end{table}

The fundamental frequencies with $m=0$ show an interesting feature:
there is a linear scaling between the real and imaginary parts of
$\omega$ and the black hole event horizon, a
characteristic first observed in~\cite{hor2000}. In that work the
temperature and the quasinormal modes are fitted by a straight line
for large black holes (high $r_+$) in AdS universes. However, for
intermediate size black holes (with size of the order of the AdS
radius) this scaling disappears. One of the reasons put forward by the
authors is that when the temperature is slowly lowered, one encounters
the Hawking-Page transition and the supergravity description is no
longer valid, {\it i.e.}, the relaxation time is not related to the
imaginary part of the fundamental quasinormal frequency anymore. In
our case we still preserve the scaling with the temperature even for
intermediate size black holes since the relation $T \propto r_+$ is always
valid. Thus, using the same argument given in Ref. \cite{hor2000} we
can conclude that this scaling is kept because there are no phase
transitions in the (2+1)-dimensional GB-BTZ black hole, a fact that we
will briefly discuss in Sec. \ref{sec5}. 

\begin{figure}[htb]
\begin{center}
\subfigure{\includegraphics[scale=0.36]{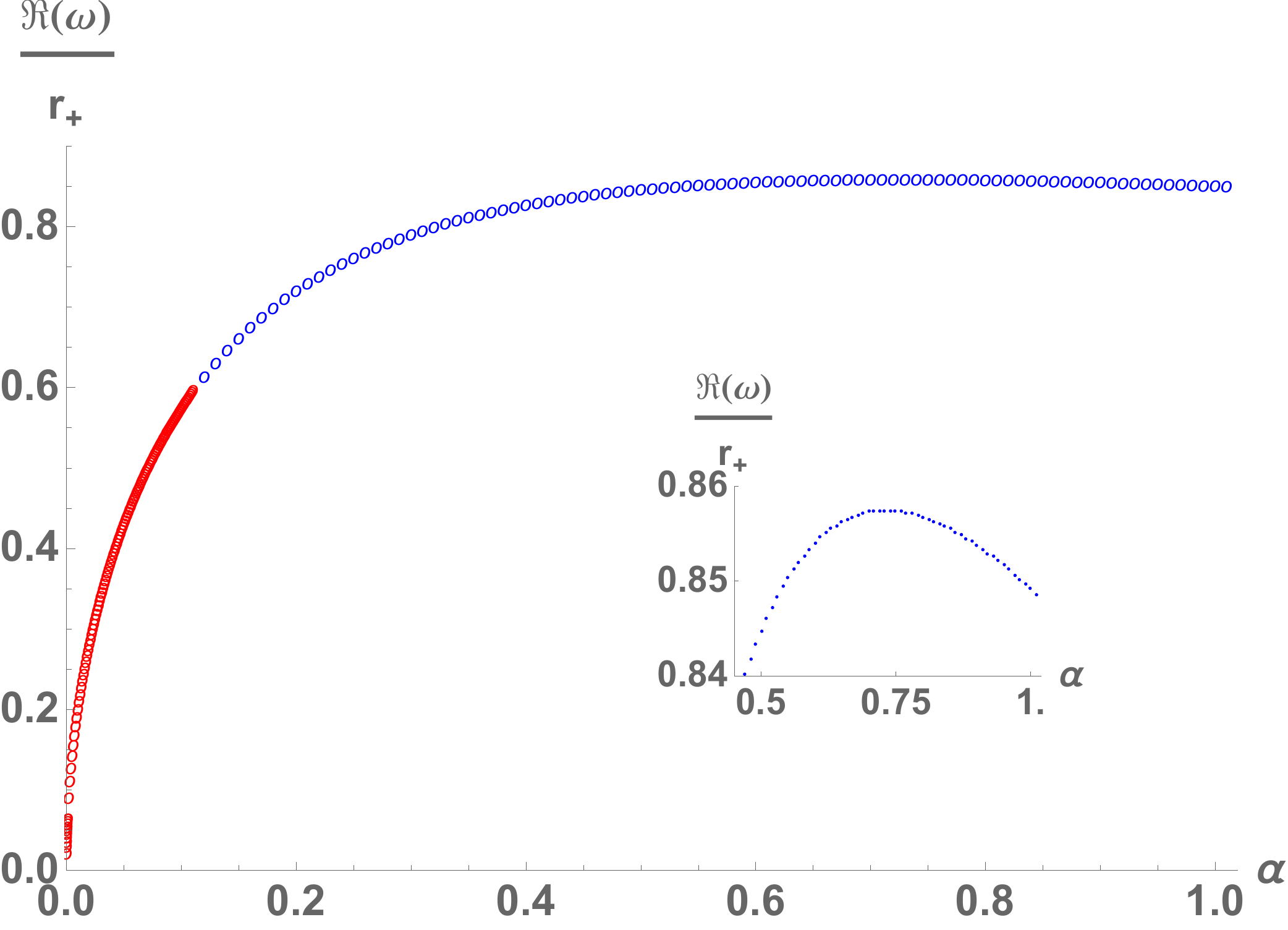}}\hskip 5ex
\subfigure{\includegraphics[scale=0.36]{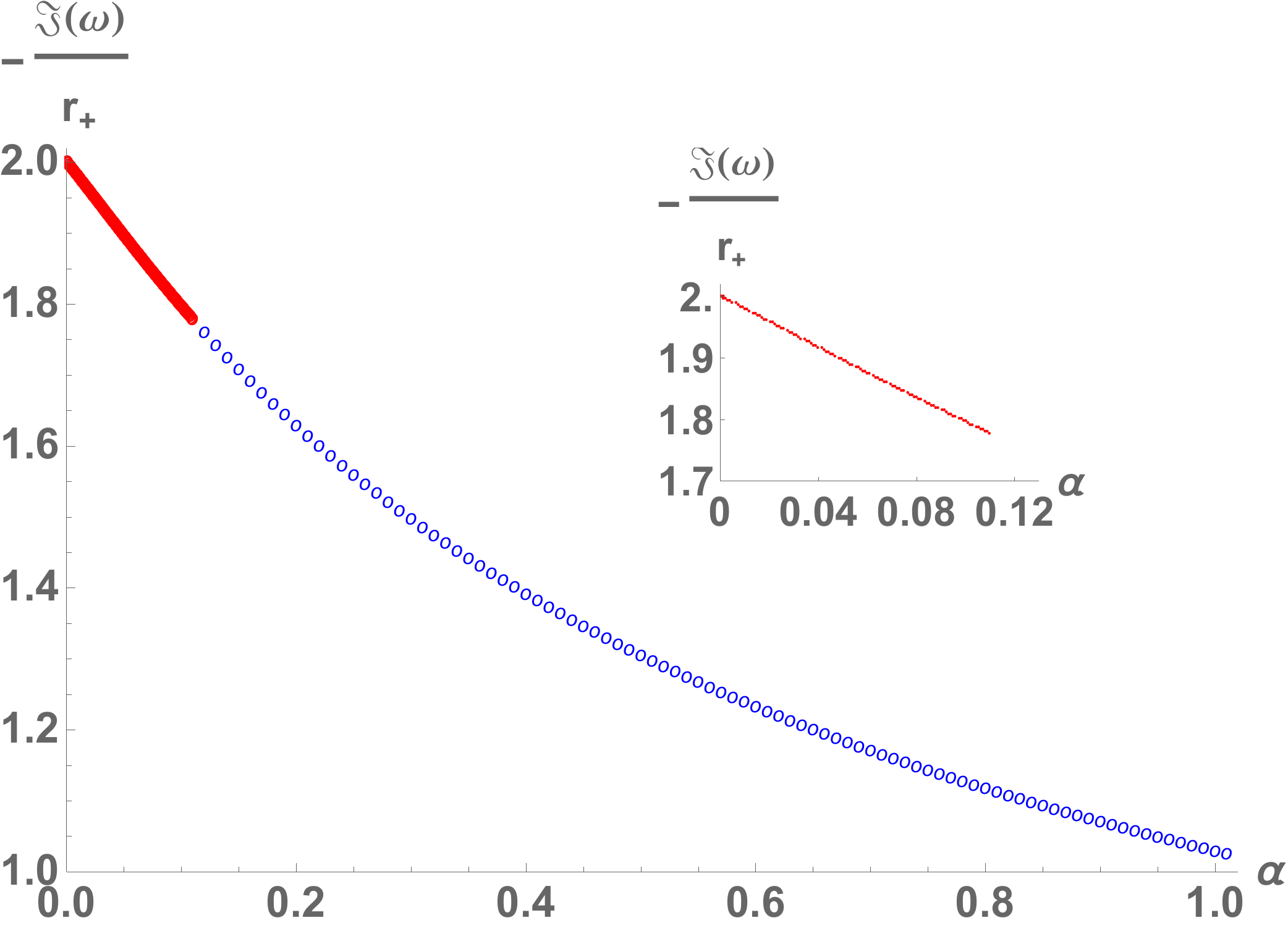}}\hskip 5ex
\caption{Quasinormal modes of the GB-BTZ black hole for different coupling parameter. The azimutal number of the field reads $m=0$.}
\label{scale}
\end{center}
\end{figure}

In Fig. \ref{scale} we show the quasinormal modes for different values
of $\alpha$ parameter. The interesting feature here is the increment
of the value of $\Re (\omega )$ with increasing $\alpha$, together
with the attenuation of $-\Im ( \omega)$. In the small-$\alpha$ regime
($m=0$) our results for the real part of the frequency suggest a
scaling given by   
\be
\label{e3}
\Re (\omega ) = (0.843-0.738 e^{-10.663 \alpha })r_+ \,,
\ee
and a linear scaling for the imaginary part expressed by
\be
\label{e4}
-\Im (\omega ) = (1.9999 - 2.0435\alpha )r_+ \,,
\ee
with linear correlation factor $R^2=0.99976$.

In Fig. \ref{scale}, we observe that the real part of the quasinormal
frequencies increases with $\alpha$, reaching a maximum at
$\alpha=\alpha_{max}$ and then starts to decrease. The same behavior
was obtained in the case of massless scalar perturbations of the
$(3+1)$-dimensional GB black hole~\cite{Konoplya:2020bxa}, where the
value $\alpha=\alpha_{max}$ indicates the possibility of gravitational
instabilities since $\Re(\omega)$ is non-monotonic. 

Since no negative potential was present in the region I of the
Penrose diagram for every tested parameter, we consistently found no
instabilities in the propagation of the scalar field subject to
well-behaved initial data. Thus, the scalar perturbation can be
decomposed after an initial burst in the traditional towers of
quasinormal modes labeled by an overtone number. Whether more than one
family of quasinormal modes can exist in the (2+1)-dimensional GB-BTZ
black hole geometry remains an open issue to be further investigated.  

In the next section we follow our stability study with the Weyl field,
whose perturbative analysis is also performed with the same tools
described in the present section.  

%------------------------------------------------------------%
\section{Probe massless spinorial field}\label{sec3}

In this section we are going to consider the problem of a massless
spinor field $\Phi$ evolving in the geometry of the (2+1)-dimensional GB-BTZ black
hole (\ref{new_bh}). The equation that dictates the dynamics of $\Phi$
is the well-known Dirac equation in its covariant form, 
\begin{equation}\label{eq_dirac}
i\gamma^{(a)}e_{(a)}^{\;\mu}\nabla_{\mu}\Phi = 0,
\end{equation}
where our index notation is the following, Latin indices
enclosed in parenthesis refer to the coordinates defined in the flat
tangent space and Greek indices indicate the spacetime coordinates. In
the tangent space we define the triad basis as in  
Eq. (\ref{triad}) and the spinor covariant derivative
$\nabla_{\mu}$ is given by the following expression,
\begin{equation}\label{cov_derivative}
\nabla_{\mu} = \partial_{\mu} +\frac{1}{8}\omega^{(a)(b)}\left[\gamma_{(a)},\gamma_{(b)}\right],
\end{equation}
in terms of spin connections $\omega_{\mu}^{(a)(b)}$ and gamma
matrices $\gamma^{(a)}$, which can be written in terms of usual Pauli
matrices \footnote{In this work we set $\gamma^{(0)}=i\sigma_2$,
  $\gamma^{(1)}=\sigma_1$ and $\gamma^{(2)}=\sigma_3$.}. The
components of the spin connection can be computed using the expression
in terms of the triad and the spacetime metric connections
$\Gamma_{\mu\rho}^{\nu}$ as 
\begin{equation}\label{spin_connections1}
\omega_{\mu}^{(a)(b)} = e_{\nu}^{(a)}\partial_{\mu}e^{(b)\nu} + e_{\nu}^{(a)}\Gamma^{\nu}_{\mu\rho}e^{\rho (b)}.
\end{equation}
The explicit expressions for the triad basis and the metric
connections are given in Appendix \ref{ap2}. Here we list the two
non-vanishing components of $\omega_{\mu}^{(a)(b)}$, computed using
the expressions (\ref{triad})-(\ref{connections}), 
\begin{equation}\label{spin_connections2}
\omega_{t}^{(t)(r)} = \frac{1}{2}\frac{df}{dr}\,, \quad
\omega_{\varphi}^{(r)(\varphi)} = -\sqrt{f}.
\end{equation}
The spinor field $\Phi$ can be written in terms of its two-components
$\Phi_1$ and $\Phi_2$ as 
\begin{equation}\label{spinor_1}
\Phi = 
\begin{pmatrix}
\Phi_1(t,r,\varphi) &  \\
\Phi_2(t,r,\varphi)&  \\ 
\end{pmatrix}.
\end{equation}
Using the tortoise coordinate $r_{*}$ and redefining the spinor components as 
\begin{equation}\label{spinor_2}
\begin{pmatrix}
\Phi_1(t,r,\varphi) &  \\
\Phi_2(t,r,\varphi)&  \\ 
\end{pmatrix}=\begin{pmatrix}
i\left(r^2 f\right)^{1/4}e^{-i\omega t +im\varphi}Y_+(r) &  \\
\left(r^2 f\right)^{1/4}e^{-i\omega t +im\varphi}Y_-(r)&  \\ 
\end{pmatrix},
\end{equation}
the Dirac equation (\ref{eq_dirac}) can be cast to the following form
\begin{equation}\label{eq_dirac_2}
\left(\frac{d}{dr_{*}}\pm i\omega \right)Y_{\pm} = WY_{\mp},
\end{equation}
where the superpotential $W$ is given by
\begin{equation}\label{superpotential}
W = m\frac{\sqrt{f}}{r}.
\end{equation}

\begin{figure}
\subfigure{\includegraphics[height = 6cm,width=8.0cm]{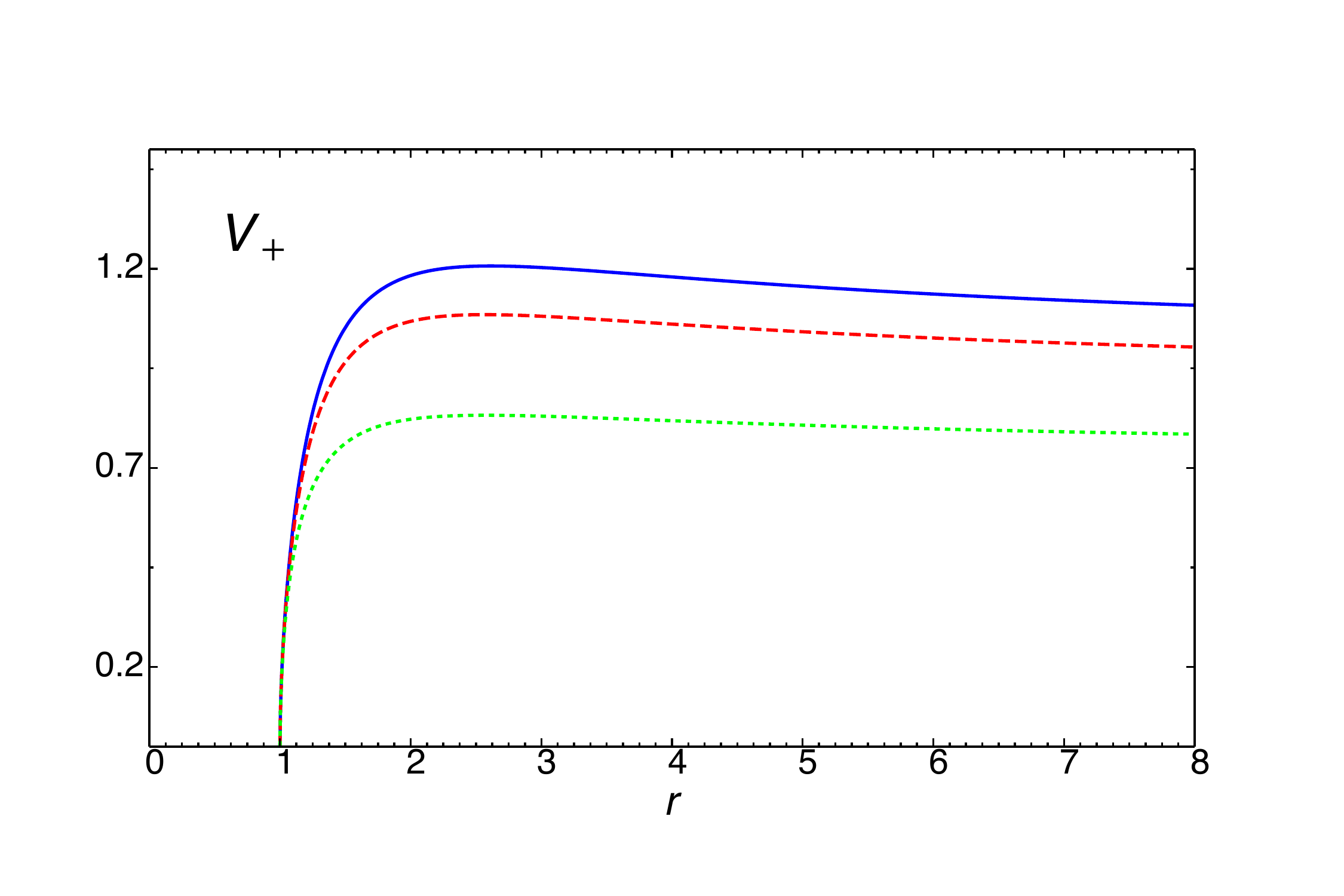}}
\subfigure{\includegraphics[height = 6cm,width=8.0cm]{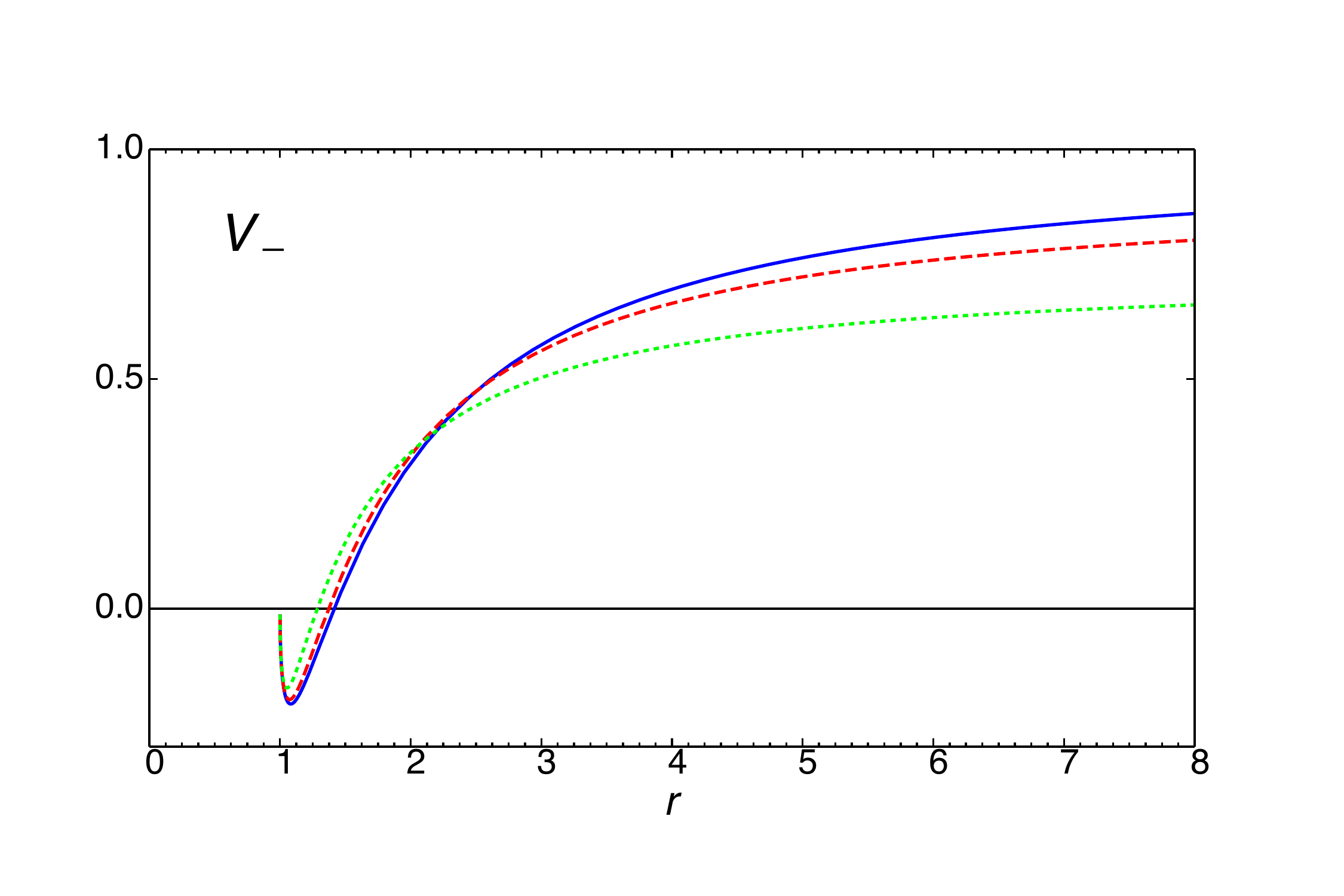}}
\caption{{\it{Left panel:}} Potential $V_{+}$ for the massless spinor with $\alpha = 0$ (blue), $\alpha = 0.1$ (dashed red), and $\alpha = 0.5$ (dotted green). {\it{Right panel:}} Potential $V_{-}$ for the massless spinor with the same values of $\alpha$ as in the left panel. }
\end{figure}

The final step to obtain the so-called superpartner potentials
$V_{\pm}$ is to introduce the
pair of coordinates $R_{+}$ and $R_{-}$ as $R_{\pm} = Y_{+}\pm Y_{-}$
into (\ref{eq_dirac_2}), so that 
\begin{equation}\label{eq_dirac_3}
\left(\frac{d^2}{dr_{*}^2}+\omega^2\right)R_{\pm} = V_{\pm}R_{\pm}\,,
\end{equation}
where $V_{\pm}$ can be expressed in terms of the superpotential $W$ as 
\begin{equation}\label{potential_spinor}
V_{\pm} = W^2 \pm \frac{dW}{dr_{*}}\,.
\end{equation}
Finally, the explicit form of the superpartner potentials for the
massless spinorial field evolving in the spacetime of the
(2+1)-dimensional GB-BTZ black hole is  
\begin{equation}\label{superpotentials}
V_{\pm} = m^2 \frac{f}{r^2} \pm m\frac{\sqrt{f}}{r}\left[\frac{1}{2}\frac{df}{dr}-\frac{f}{r}\right].
\end{equation}

The propagation of a massless spinor field in the black hole geometry
dictated by Eq. (\ref{eq_dirac_3}) recovers the
quasinormal modes through the signal of field profiles. The profiles
are obtained as described in the previous section with the double null
integration technique. After an initial perturbation (gauge), the quasinormal
evolution takes place and the frequencies are drawn with the
Prony method, mentioned in the previous section. We use the usual
gaussian packages in null coordinates as initial surface
to evolve the field.

\begin{table}
  \centering
 \caption{The fundamental quasinormal modes for a massless spinorial field with $L=1$ and azimutal number $m = 1$ with potential $V_+$. The frequencies represent a stable field evolution with a purely imaginary decay, except for $r_+=1$.}
\addtolength\tabcolsep{6pt}
    \begin{tabular}{c|ccccccc}
    \hline
$r_+$	 & \multicolumn{7}{c}{$\alpha$} \\
	&	0.001	&	0.01	&	0.1	&	0.2	&	0.3	&	0.4	&	0.5	\\
\hline \hline
1	&	1.10795		&	1.10722		&	1.09530		&	1.07726		&	1.05788		&	1.03867		&		1.02019	\\
	&	-0.86224i		&	-0.85258i		&	-0.76950i		&	-0.69847i	&	-0.64240i		&	-0.59673i		&	-0.55863i	\\
\hline
5	&	-2.71475i		&	-2.71715i		&	-2.74155i		&	-2.77000i	&	-2.80042i	&	-2.83343i	&	-2.86972i \\
\hline
10	&	-5.10072i		&	-5.10171i		&	-5.11161i		&	-5.12270i	&	-5.13402i	&	-5.14566i	&	-5.15771i \\
\hline
50	&	-25.01925i		&	-25.01943i		&	-25.02117i		&	-25.02308i	&	-25.02501i	&	-25.02695i	&	-25.02893i \\
\hline
100	&	-50.00967i		&	-50.00976i		&	-50.01062i		&	-50.01158i	&	-50.01253i	&	-50.01350i	&	-50.01448i \\	
    \hline  
    \end{tabular}
  \label{tb3}
\end{table}

\begin{table}
  \centering
 \caption{The fundamental quasinormal modes for a massless spinorial field with $L=1$ and azimutal number $m = 1$ with potential $V_-$. The frequencies represent a stable field evolution with a purely imaginary decay.}
\addtolength\tabcolsep{6pt}
    \begin{tabular}{c|ccccccc}
    \hline
$r_+$	 & \multicolumn{7}{c}{$\alpha$} \\
	&	0.001	&	0.01	&	0.1	&	0.2	&	0.3	&	0.4	&	0.5	\\
\hline \hline
1	& -0.10731i	&	-0.10670i	&	-0.10115i	&	-0.095998i	&	-0.091609i	&	-0.087790i	&	-0.084415i	\\
5	& -2.33789i	&	-2.33677i	&	-2.32594i	&	-2.31449i	&	-2.30348i	&	-2.29281i	&	-2.28241i	\\
10	& -4.91111i	&	-4.91040i	&	-4.90340i	&	-4.89582i	&	-4.88837i	&	-4.88098i	&	-4.87362i	\\
50	& -24.98117i	&	-24.98100i	&	-24.97936i	&	-24.97756i	&	-24.97576i	&	-24.97395i	&	-24.97211i	\\
100	& -49.99043i	&	-49.99035i	&	-49.98951i	&	-49.98858i	&	-49.98766i	&	-49.98673i	&	-49.98579i	\\
    \hline  
    \end{tabular}
  \label{tb4}
\end{table}

The quasinormal modes are listed in the Tables \ref{tb3} and
\ref{tb4}. There we can verify an interesting behavior: for
increasing $\alpha$ the damping factor varies in opposite directions,
increasing for $V_+$ and decreasing for $V_-$. This effect is more
pronounced for small $r_+$ such that the spectra of larger black holes
are very mildly influenced by the variation of $\alpha$. A scaling
between $r_+$ and the quasinormal frequency emerges for large black
holes, 
\be
\label{sc1}
\omega \simeq \frac{r_+}{2}
\ee 
for both $V_+$ and $V_-$. This is the same result obtained for the
quasinormal modes of the BTZ black hole to zeroth order in
$m$~\cite{deOliveira:2018weu}. Interestingly, $\alpha$ plays no
important role in the spectra for large black holes in the spinorial field
profile contrarily to the scalar case. 

The evolution of the massless spinorial field was extensively
investigated with our methods and found to be stable. The field profile,
after an initial burst, decomposes into a tower of quasinormal modes from
which specific cases are listed in Tables \ref{tb3} and
\ref{tb4}. This is an expected result for the potential $V_+$,
however, it is not granted for the potential $V_-$ since a small region with
$V<0$ exists for $r>r_+$ in this case. 

As usual, in AdS-like black holes the spectra for both $V_+$ and
$V_-$ are not the same. Such behavior of isospectrality of the
potentials is found whenever a series expansion of the transmission
coefficients associated to $W$ is the same for $V_+$ and
$V_-$~\cite{Chandrasekhar:1985kt,cardoso2004quasinormal}. The fact that
\be
\label{sc2}
W\Big|_{r\rightarrow r_+}^{r \rightarrow \infty} = W_\infty > 0\,.
\ee 
is sufficient to break the isospectrality of the potentials.

\section{Scalar quasinormal modes in the hydrodynamical
  approximation}\label{sec4} 

In this section we are going to consider the hydrodynamical limit of
probe scalar field. In general, an interacting theory can be described
by means of hydrodynamics in the limit of large wavelength and small
wavenumbers compared to the typical temperature of the
system~\cite{Landau1987FluidMechanics}. From Gauge/Gravity
correspondence it is well-know that the characteristic thermalization
timescale for a dual thermal state at the boundary is given by the
inverse of the imaginary part of the fundamental quasinormal frequency
in the hydrodynamical limit~\cite{hor2000}. Such a result has been
confirmed in $(2+1)-$dimensional black holes with Lifshitz
scaling~\cite{Abdalla2012ThreeDimensional}.

In order to establish this limit, we define the quantities
$\mathfrak{w}=\omega/2\pi T$ and $\mathfrak{q} = m/2\pi T$ and
consider the limit $\mathfrak{q}\rightarrow 0$, such that the radial
equation for the massless scalar field can be cast to 

\begin{equation}\label{kg_hidro1}
R''(u) +\left[\frac{h'}{h} -\frac{1}{u}\right]R'(u) +\frac{4\alpha^2}{h^{2}L^4}\mathfrak{w}^2 R(u) = 0,
\end{equation}
where we have performed the change of variable $u=r_{+}/r$ and defined
\begin{equation}\label{hfunction}
h = 1-\left[1+\frac{4\alpha}{L^2}(1-u^2)\right]^{1/2}.
\end{equation}
In the case in which $\mathfrak{w}<<1$ we expand $R(u)$ in powers of $\mathfrak{w}$,
\begin{equation}\label{expansion1}
R(u)\approx h(u)^{\sigma}\left(F_{0}(u) + i\mathfrak{w}F_{1}(u) +\mathcal{O}(\mathfrak{w}^2)\right).
\end{equation}
The exponent $\sigma$ is determined by imposing the ingoing boundary
condition for the scalar field at the location of the black hole event
horizon $r_{+}$. We thus obtain $\sigma = -i\mathfrak{w}/2$.

Substituting the expansion (\ref{expansion1}) in the scalar field
radial equation (\ref{kg_hidro1}) we obtain two ordinary differential
equations for the functions $F_{0}(u)$ and $F_{1}(u)$, 
\begin{eqnarray}\label{F0}
F_{0}''(u)+\left(\frac{h'}{h} - \frac{1}{u}\right)F_{0}'(u)=0\,,\\ \label{F1}
F_{1}''(u) +\left(\frac{h'}{h} - \frac{1}{u}\right)F_{1}'(u) -
\frac{h'}{h} F_0'+\left(\frac{h'}{u}-h''\right)\frac{F_0}{2 h}= 0\,.
\end{eqnarray}
In order to analyze the influence of GB coupling constant $\alpha$ on the
frequencies in the hydrodynamical limit, we will consider the
small-$\alpha$ limit. Expanding $h(u)$ in such a limit we obtain,   
\begin{equation}\label{expand_h}
h(u) \approx \frac{2}{L^2}(u^2 - 1)\left[\alpha
  +\frac{\alpha^2}{L^2}(u^2-1)\right]\,. 
\end{equation}
In this scenario the solution for Eq. (\ref{F0}) is
\begin{equation}\label{sol_F0}
F_{0}(u) = A - \frac{B}{2L^2}\ln \left[{\frac{(u^2-1)}{(L^2 +\alpha(u^2 -1))}}\right]\,,
\end{equation}
where $A$ and $B$ are constants.
To satisfy the ingoing boundary condition at the event horizon and avoid
divergences in $F_{0}(u)$, we need impose $B=0$ in (\ref{sol_F0}). Thus,
the solution becomes $F_{0}(u) = A$. With this result we solve
Eq. (\ref{F1}) for $F_1(u)$ obtaining, 
\begin{equation}
F_{1}(u) = C +\frac{2A\alpha - D}{2L^2}\ln{(u^2-1)} +\frac{D+2A(L^2-\alpha)}{2L^2}\ln{[L^2 + (u^2 -1)\alpha]},
\end{equation}
where $C$ and $D$ are constants. Again, the solution has to be finite
as $u\rightarrow 1$, thus, we must have $D=2A\alpha$. Also, the ingoing boundary condition at the event horizon implies that $F_{1}(1) = 0$, then, 
\begin{equation}
C = -A\ln{(L^2)}.
\end{equation}
Finally, the solution for $F_{1}(u)$, finite and obeying the physical
boundary condition at the event horizon, turns to be
\begin{equation}\label{F1_2}
F_{1}(u) = A\ln{\left[1+\frac{\alpha}{L^2}(u^2 - 1)\right]}.
\end{equation}

Replacing the solutions for $F_{0}(u)$ and $F_{1}(u)$ back in
Eq. (\ref{expansion1}) we have 
\begin{equation}\label{R_2}
R(u) = A\, h(u)^{-i\mathfrak{w}/2}\left\{1+i\mathfrak{w}\ln{\left[1+\frac{\alpha}{L^2}(u^2 - 1)\right]}\right\}\,.
\end{equation}
Imposing the Dirichlet boundary condition at spatial infinity for the
scalar field, $R(0) = 0$, we arrive to the following allowed set of
frequencies,  
\begin{equation}\label{omega}
\mathfrak{w} = \frac{i}{\ln{\left(1-\frac{\alpha}{L^2}\right)}},
\end{equation}
which in terms of black hole temperature $T$ reads
\begin{equation}\label{omega2}
\omega = \frac{2\pi T i}{\ln{\left(1-\frac{\alpha}{L^2}\right)}}.
\end{equation}
The hydrodynamical frequencies are purely imaginary, showing the same
behavior as the three-dimensional black holes with Lifshitz
symmetry~\cite{Abdalla2012ThreeDimensional} and those surrounded by
anisotropic fluids~\cite{deOliveira:2018weu}. In terms of Gauge/Gravity
correspondence the perturbation of a black hole in the gravity side is
equivalent to perturb a thermal state in the gauge theory side. In
this context, the inverse of the imaginary part of the fundamental
quasinormal frequency corresponds to the relaxation time which the
perturbed thermal state needs in order to return to thermal
equilibrium. Thus, in our case this timescale is given by
\begin{equation}\label{timescale}
\tau = \frac{\ln{\left(1-\frac{\alpha}{L^2}\right)}}{2\pi T}.
\end{equation}
At high temperatures the timescale $\tau$ approaches zero
suggesting that the perturbations of thermal states in the
$(1+1)-$field theory are not long-lived. However, as $\alpha$
increases (provided that $-L^2/4<\alpha<0$) with fixed
black hole temperature, the timescale $\tau$ increases as well
indicating the possibility of having long-lived perturbations in the gauge
theory. 

In the next section we will discuss some aspects of the thermodynamics
of (2+1)-dimensional GB-BTZ black holes.

%%%%%%%%%%%%%%%%%%%%%%%%%%%%%%%%%%%%%%%%%%%%
\section{Thermodynamical aspects}\label{sec5}

The thermodynamics of the black hole described by the negative branch
of Eq. (\ref{new_bh}) is very simple as quoted
by~\cite{Hennigar:2020fkv}, in which the main thermodynamical variables
are listed as follows,
\begin{eqnarray}\label{thq}
&T = \frac{r_{+}}{2\pi L^2}\,, \quad P=\frac{1}{8\pi L^2}\,, \quad
V=\pi r_+ ^2 \,,& \nonumber \\
&M = \frac{r_+ ^2}{8L^2}\,, \quad S = \frac{\pi r_+}{2} \,, \psi_\alpha
= 0\,,&
\end{eqnarray}
where $\psi_\alpha$ is the potential conjugated to the GB parameter. 
In particular, we notice that its Hawking temperature grows
monotonically with $r_+$, so that there are no phase transitions. This
fact can also be seen from the simple equation of state obtained
combining $T$, $V$, and $P$ in the list of Eqs.(\ref{thq}),
\begin{equation}
P = \frac{T}{v}\,,
\end{equation}
where we have defined the specific volume as ${v} =
4\sqrt{V/\pi}$. This equation clearly has no critical points.

Another analysis that supports this conclusion is the study of null
geodesics in this geometry. It is known that the photon sphere radius
and the impact parameter related to it play an interesting role during
a black hole phase transition and can serve as order parameters to
describe such a phenomenon~\cite{PhysRevD.97.104027}. Thus, by
considering the Lagrangian
\begin{equation}
2{\cal L} = -f(r) \dot t^2 + \frac{\dot r^2}{f(r)} + r^2 \dot\varphi^2 \,,
\end{equation}
and the constants of motion defined by the generalized momenta
corresponding  to $t$ and $\varphi$,
\begin{equation}
p_t = -f(r) \dot t = -E \,,\qquad p_\varphi = r^2 \dot\varphi = L\,,
\end{equation}
we can obtain the radial equation for a photon moving in this
spacetime,
\begin{equation}
\dot r^2 + V_{eff} = 0 \,, \quad \hbox{with }\quad V_{eff} =
\frac{fL^2}{r^2} - E^2 \,.
\end{equation}
Applying the usual conditions in order to obtain the photon
sphere\footnote{As we are working in (2+1) dimensions, {\it photon
  circumference} would be a more appropiate term.} radius $r_{ps}$
\begin{equation}\label{pscond}
V_{eff}=0\,, \quad \frac{dV_{eff}}{dr} = 0 \,, \quad \frac{d^2
  V_{eff}}{dr^2} <0 \quad \hbox{at } r_{ps} \,,
\end{equation}
we notice that the second equation of set (\ref{pscond}) cannot be
solved for any finite $r_{ps}$. In fact, as happens in BTZ solution,
this (2+1)-dimensional GB-BTZ black hole has no photon circumference. Then, the
absence of phase transitions becomes evident. 

In what follows we discuss some entropy aspects for this geometry,
namely, we calculate the Bekenstein entropy bound and the
leading correction to the black hole entropy using the brickwall
method. 

\subsection{Entropy bound}

We consider the motion of a particle near a black hole described by
the metric (\ref{new_bh}). The constants of motion correspond to the
energy and angular momentum of the particle, respectively,
\begin{eqnarray}\label{EJ}
E &=& \pi_t = g_{tt} \dot t \,, \nonumber \\
J &=& -\pi_\varphi = -g_{\varphi\varphi} \dot \varphi \,.
\end{eqnarray}
In addition, the energy conservation for a particle of mass $m$
implies,
\begin{equation}\label{Econs}
-m^2 = g^{\mu\nu}\pi_\mu \pi_\nu \,,
\end{equation}
so that 
\begin{equation}\label{Eeq}
r^2 E^2 - J^2 f(r) - m^2 r^2 f(r) = 0 \,,
\end{equation}
whose solution gives an expression for the particle's energy,
\begin{equation}\label{ener}
E = \frac{\sqrt{f(r)}}{r} \sqrt{J^2 + m^2 r^2} \,.
\end{equation}

As the particle is gradually approaching the black hole, it finally
reaches the event horizon when the proper distance from its center of mass
to this horizon equals $R$, the characteristic dimension of the
particle, 
\begin{equation}\label{Rdist}
\int_{r_+} ^{r_+ + \delta(R)} \sqrt{g_{rr}} dr = R \,,
\end{equation}
where the upper limit of the integral represents the point of capture
of the particle by the black hole. Expanding to first order we obtain
for $\delta$, 
\begin{equation}\label{deltaR}
\delta (R) \approx \frac{r_+ R^2}{2L^2} \,.
\end{equation}
And we can minimize the energy (\ref{ener}) at the point of capture with respect
to the particle's angular momentum, {\it i.e.},
\begin{equation}\label{minE}
\frac{dE}{dJ}\bigg|_{r_+ +\delta} = 0 \quad \Rightarrow \quad J=0\,,
\end{equation}
thus obtaining,
\begin{equation}\label{Emini}
E_{min} = m \sqrt{f(r_+ + \delta)} = \frac{mr_+ R}{L^2}\,.
\end{equation}
Now, according to the first law of thermodynamics we have that 
\begin{equation}\label{1law}
dM = E_{min} = T\,dS = \frac{\kappa}{2\pi} dS \,,
\end{equation}
being $\kappa$ the surface gravity at the event horizon,
\begin{equation}\label{sgrav}
\kappa = \frac{f'}{2}\bigg|_{r=r_+} = \frac{r_+}{L^2}\,.
\end{equation} 
At the same time, the generalized second law of thermodynamics says
that after the capture of the particle the entropy of the black hole
cannot decrease, 
\begin{equation}\label{2law}
S_{BH} (M+dM) \geq S_{BH} (M) + S \,.
\end{equation}
Thus, combining both laws we can obtain an upper bound on the entropy
of the particle 
\begin{equation}
S \leq dS = S_{BH} (M+dM) - S_{BH} (M) = 2\pi m R \equiv 2\pi ER \,.
\end{equation}
This bound shows to be independent of the black hole parameters and
agrees with the universal result obtained by
Bekenstein~\cite{PhysRevD.23.287}, valid for any dimensionality.

\subsection{Entropy semiclassical correction}

In order to find semiclassical corrections to the black hole entropy,
we use 't Hooft's brickwall method~\cite{THOOFT1985727}. This method
considers a thermal bath of scalar fields quantized using the
partition function of statistical mechanics, whose leading
contribution yields the Bekenstein-Hawking formula. The method
introduces certain conditions on the scalar field $\Phi$ aiming to avoid
divergences, namely, an ultraviolet cut-off near the event horizon
($\Phi=0$ for $r\leq r_+ + \epsilon$) and an infrared cut-off far from
the black hole ($\Phi=0$ for $r \geq L \gg r_+$). 

The scalar field of mass $\mu$ obeys the massive version of the Klein-Gordon
equation given by Eq. (\ref{kg1}),
\begin{equation}\label{kgm}
\frac{1}{\sqrt{-g}} \partial_\mu (\sqrt{-g} g^{\mu\nu} \partial_\nu
\Phi) -\mu^2 \Phi = 0\,.
\end{equation}
Using the {\it ansatz} $\Phi (t,r,\varphi) = e^{-iEt+im\varphi} R(r)$ the radial
part of Eq. (\ref{kgm}) turns out to be
\begin{equation}\label{radeq}
\frac{d^2 R}{dr^2} + \left( \frac{f'}{f} + \frac{1}{r} \right)
\frac{dR}{dr} + \frac{1}{f} \left( \frac{E^2}{f}-\frac{m^2}{r^2}
-\mu^2 \right) R = 0 \,.
\end{equation}
In order to obtain the radial wave number $K$, we use a WKB
approximation for $R(r) \sim e^{iS(r)}$, with $S(r)$ being a rapidly varying
phase. To leading order the only significative contribution to the
radial wave number comes from the first derivative of $S$ obtained 
from the real part of Eq. (\ref{radeq}),
\begin{equation}\label{waven}
K \equiv S' = \frac{1}{\sqrt{f}} \left[ \frac{E^2}{f} - \left(
  \frac{m^2}{r^2} +\mu^2 \right) \right]^{1/2}\,.
\end{equation}
Then, we use $K$ to quantize the number of radial modes $n_r$ of the
field as follows,
\begin{equation}\label{radmodes}
\pi n_r = \int_{r_+ +\epsilon} ^L K(r,m,E)\, dr \,. 
\end{equation}

Moreover, in order to find the black hole entropy of the system, we
calculate the Helmholtz free energy $F$ of the scalar thermal bath
with temperature $\beta^{-1} = \kappa/2\pi$ as follows,
\begin{equation}
F = \frac{1}{\beta} \int 2\,dm \int \ln(1-e^{-\beta E})\, dn_r = -\int 2
\,dm \int \frac{n_r}{e^{\beta E}-1} \,dE \,.
\end{equation}
Performing the integral in $m$ and using Eq. (\ref{radmodes}) we obtain
\begin{equation}
F = -\frac{1}{2} \int_0 ^\infty \frac{dE}{e^{\beta E}-1} \int_{1
  +\bar\epsilon} ^{\bar L} \frac{r_+ ^2 y}{\sqrt{f(y)}} \left(
\frac{E^2}{f(y)} - \mu^2 \right) dy \,,
\end{equation}
where we rescaled the quantities, $y=r/r_+$, $\bar L = L/r_+$, and
$\bar\epsilon = \epsilon/r_+$. Thus, the metric coefficient can be
written as
\begin{equation}
f(y) = -\frac{r_+ ^2}{2\alpha} y^2 \left[ 1-\sqrt{1+
    \frac{4\alpha}{y^2 L^2}(y^2-1)} \right] \,.
\end{equation}
Expanding near the event horizon where $y\rightarrow 1$ and performing
the Bose-Einstein integral we get
\begin{equation}\label{fhelm}
F \approx -\frac{\zeta (3)}{\beta^3} \frac{(2\alpha)^{3/2}}{r_+}
\int_{1 + \bar\epsilon} ^{\bar L} \left[ -1 +
  \sqrt{1+\frac{4\alpha}{L^2} (y^2-1)} \right]^{-3/2} dy \,,
\end{equation}
with $\zeta(x)$ being the Riemann zeta function. 
The semiclassical correction we are searching comes from the divergent
contribution of Eq. (\ref{fhelm}), {\it i.e.}, from the lower limit of
the integral, whose leading order term reads,
\begin{equation}
F_\epsilon = -\frac{\zeta (3)L^3}{\beta^3 \sqrt{2r_+ \epsilon}}\,.
\end{equation}

The corresponding entropy $S_\epsilon $ follows directly,
\begin{equation}\label{entropy1}
S_\epsilon = \beta^2 \frac{\partial
F_\epsilon}{\partial\beta} = \frac{3\zeta(3) L^3}{\beta^2 \sqrt{2r_+ \epsilon}}\,.
\end{equation}
In order to write this correction in a more familiar way, we use the
proper thickness $\xi$ defined as
\begin{equation}
\xi = \int_{r_+} ^{r_+ + \epsilon} \sqrt{g_{rr}}\,dr \approx \frac{L\sqrt{2\epsilon}}{\sqrt{r_+}}\,,
\end{equation}
as well as the event horizon ``area'' $A=2\pi r_+$ and the Hawking
temperature $T = \frac{1}{\beta} = \frac{r_+}{2\pi L^2}$, obtained
from the surface gravity (\ref{sgrav}), to finally achieve,
\begin{equation}\label{scent}
S_\epsilon = \frac{3 \zeta(3) A}{8\pi^3 \xi}\,,
\end{equation}
which is a universal expression in three-dimensional
gravity~\cite{ART001199184,Kim2009ThermodynamicsOW}.

%------------------------------------------------------------%
\section{Final Remarks}\label{sec6}

In this paper, we have studied the perturbative and thermodynamical
aspects of the $(2+1)$-dimensional GB-BTZ black hole found by Hennigar
{\it{et. al.}}~\cite{Hennigar:2020fkv,Hennigar:2020drx}. This solution
describes a family of lower-dimensional black holes parametrized by
the mass term $M$, the AdS$_3$ radius $L$, and the GB coupling
constant  $\alpha$. Also, the BTZ limit of the solutions exists as
$\alpha\rightarrow 0$, and the event horizon is located at
$r_{+}=LM^{1/2}$. In order to understand the role of the GB coupling
constant $\alpha$ in the context of the black hole stability problem,
we performed the computations and analysis of the GB-BTZ black hole
quasinormal spectrum. In addition, the Bekenstein entropy bound and
the semiclassical correction to the Bekenstein-Hawking entropy were
also computed.  

We analyzed two different types of perturbations represented by a scalar
field and a massless spinorial field. For intermediate
black holes, both scalar and spinorial perturbations are affected
reasonably by the variation of the GB coupling constant, although the
influence in the scalar case is much more pronounced. This is also
true for large black holes perturbed by the scalar field. Interestingly enough,
such a picture changes for large black holes in the massless
spinorial case, where the influence of the coupling constant is almost
insignificant. To first order in the angular momentum we can
understand the perturbation in a common ground as the same reported
for a BTZ black hole when $\alpha=0$ (see
e. g.~\cite{deOliveira:2018weu}), establishing no role played by $\alpha$
on the perturbation. In both cases analyzed here the extensive search
for profiles with different geometry parameters results in a stable
spacetime against the field perturbations. 

The quasinormal modes obtained for the scalar and spinorial
perturbations in the background of the GB-BTZ black hole are assembled
in Tables \ref{tb1} to \ref{tb4} and display interesting features of
the geometry, already described in the previous sections. In the
scalar case, for instance, we remark the presence of a peak 
in the graphic of $\frac{\Re (\omega )}{r_+}$ vs. $\alpha$ and the
linear scaling of $\alpha $ and $-\frac{\Im (\omega )}{r_+}$ in the
small-$\alpha$ regime. Moreover, for the fundamental mode we see a
linear scaling between the quasinormal frequencies and the temperature
of the black hole, which can be interpreted as an absence of phase
transitions in the model, also confirmed by the thermodynamical
analysis. 

As for the massless spinorial perturbation, only oscillatory modes are
present for the scalar field with $V_-$ potential, different from
what is found in the pure BTZ case~\cite{Cardoso:2001hn} and the
(2+1)-dimensional Lifshitz black hole~\cite{Cuadros_Melgar_2012}.  
In the case of $V_+$ potential a critical $r_+$ exists such that it
points out the transition from oscillatory to non-oscillatory modes, a
behavior also found in $(2+1)$-dimensional-black holes with anisotropic
fluids~\cite{deOliveira:2018weu}.  

Subsequently, regarding the quasinormal spectrum due to scalar perturbations,
we found that the hydrodynamical or high-temperature
approximation leads to an exact result for the quasinormal spectrum 
$w= i(2\pi T)/\ln(1-\alpha/L^2)$ in the small-$\alpha$ limit provided
that $-L^2/4<\alpha<0$. As the hydrodynamical
frequencies are purely imaginary, there is not oscillatory phase in
this limit either. In the context of Gauge/Gravity correspondence this
result suggests that perturbations of thermal states in the
$(1+1)-$field theory are not long-lived. 

Afterwards, we briefly discussed the thermodynamics of the
(2+1)-dimensional GB-BTZ black hole, stressing that there are no phase
transitions since its temperature is a monotonic growing function, a
result that is also reinforced by the absence of a photon
circumference in the geometry. Furthermore, we calculated the
Bekenstein entropy bound for an object captured by this black hole
obeying the first and second laws of thermodynamics. Our result
complies with the universality of the bound. In addition, we computed
the leading semiclassical correction to the Bekenstein-Hawking entropy
by means of the brickwall method. This correction shows a perfect
agreement with other (2+1)-dimensional black holes. 

Finally, according to our results we can conclude that the
(2+1)-dimensional GB-BTZ black holes are dynamically stable under
scalar and spinorial linear perturbations. We should also stress that
as in this dimensionality the metric or gravitational perturbations
reduce to a scalar mode only because there are no propagating 
degrees of freedom~\cite{Carlip_2005}, what also happens even in
higher-dimensional braneworld models~\cite{Cuadros_Melgar_2011}, our
stability analysis is a good candidate for a definitive answer on this
matter. Moreover, this dynamical stability is also accompanied by a
thermodynamical stability and a full agreement of our results with the
universality of entropy aspects discussed here. The stability analysis
of more general solutions, including charge or angular momentum as shown in
Ref.\cite{Hennigar_2020}, is left for future works.

%------------------------------------------------------------%
\begin{acknowledgments}
This work was partially supported by UFMT (Universidade Federal de Mato
Grosso) under grant 001/2021 - Edital PROPEQ de Apoio à Pesquisa. 
\end{acknowledgments}

\appendix

\section{Frobenius method for the Klein-Gordon equation}\label{ap1}

The Frobenius method we developed for the scalar field propagation
starts by taking Eq. (\ref{kg2}) and defining a new variable
$x=r^{-1}$. In such coordinates the field equation reads 
\be
\label{ea1}
s\psi'' +\tau \psi' + u \psi =0\,,
\ee
in which prime denotes a derivative with respect to $x$ and the
functions $s$, $\tau$, and $u$ are given by
\be
\nonumber
s=f^2x^4\equiv \sum_{n=0}^{\infty}s_n (x-x_+)^{n+\delta}\,, \\
\nonumber
\tau=fx^2(2xf+x^2f')\equiv \sum_{n=0}^{\infty}\tau_n (x-x_+)^{n+\delta}\,, \\
\label{ea2}
u=\omega^2-V\equiv \sum_{n=0}^{\infty}u_n (x-x_+)^{n+\delta}\,.
\ee
Now, using the Ansatz $\psi = \sum a_n (x-x_+)^{n+\delta}$ the solution to leading order (indicial relation) is given by the expression,
\be
\label{ea3}
\delta = \pm i\frac{\omega r_+}{2M}\,,
\ee
being the negative sign the correct one according to the right
boundary condition. Substituting the {\it ansatz} in the field equation we still retain the recurrence relation,
\be
\label{ea4}
a_n = -\frac{1}{D_n}\sum_{k=0}^{n-1}\{s_{n-k+2}[k(k-1)+\delta (2k-1)+\delta^2]+\tau_{n-k+1}[k+\delta]+u_{n-k}\} a_k
\ee
with $D_n=n(n+2\delta) s_2$. Such a expression allows us to solve the
quasinormal problem in a similar way as that described in~\cite{hor2000}.

\section{Metric connections and triad basis}\label{ap2}
The components of triad basis for the metric (\ref{new_bh}) are given by
\begin{equation}\label{triad}
e_{t}^{(a)}=\sqrt{f(r)}\,\delta_{t}^{(a)}, \hspace{0.3cm} e_{r}^{(a)} = \frac{1}{\sqrt{f}}\,\delta^{(a)}_{r},\hspace{0.3cm} e_{\varphi}^{(a)} = r\,\delta_\varphi^{(a)},
\end{equation}
and the metric connections read
\begin{eqnarray}\label{connections}\nonumber
&\Gamma_{tr}^{t} = \frac{d}{dr}\left[\ln(\sqrt{f})\right], \hspace{0.3cm} \Gamma_{rr}^{r} = \frac{d}{dr}\left[\ln\left(\frac{1}{\sqrt{f}}\right)\right],\hspace{0.3cm} \Gamma_{tt}^{r} = \frac{f}{2}\frac{df}{dr}\,,&\\
&\Gamma_{\varphi\varphi}^{r} = -rf,\hspace{0.3cm} \Gamma_{r\varphi}^{\varphi} = \frac{1}{r} \,.&
\end{eqnarray}

%------------------------------------------------------------%
\section*{References}

\bibliography{referencias}

\end{document}